\documentclass[aps,prl,twocolumn]{revtex4}  

\usepackage{bbm,graphicx}   

\def\ave#1{\langle #1 \rangle}
\def\ii{{\rm i}}

\def\tit#1{{\em #1},}
\def\etal#1{#1}

\begin{document}

\title{Spin transport in a one-dimensional anisotropic Heisenberg model}

\author{Marko \v Znidari\v c}
\affiliation{
Instituto de Ciencias F\' isicas, Universidad Nacional Aut\' onoma de M\' exico, Cuernavaca, Mexico, and\\
Faculty of Mathematics and Physics, University of Ljubljana, Ljubljana, Slovenia}

\date{\today}

\begin{abstract}
We analytically and numerically study spin transport in a one-dimensional Heisenberg model in linear-response regime at infinite temperature. It is shown that as the anisotropy parameter $\Delta$ is varied spin transport changes from ballistic for $\Delta <1$ to anomalous at the isotropic point $\Delta=1$, to diffusive for finite $\Delta > 1$, ending up as a perfect isolator in the Ising limit of infinite $\Delta$. Using perturbation theory for large $\Delta$ a quantitative prediction is made for the dependence of diffusion constant on $\Delta$.  
\end{abstract}


\maketitle
The one-dimensional spin $1/2$ Heisenberg model is one of the oldest quantum models~\cite{Heisenberg:28} being also the simplest model of interacting quantum particles. Despite being exactly solvable by the Bethe ansatz~\cite{Bethe:31}, calculating its dynamical properties, like transport, is by no means simple. Understanding transport in the Heisenberg model is important for several reasons. First, it is still not known what are the necessary requirements for a system to display phenomenological transport laws where a current is proportional to the gradient of the driving field. Second motivation comes from the condensed matter where one would like to understand transport in strongly-correlated electron systems. A paradigmatic model is the Hubbard model, thought to be related to the problem of high-$T_{\rm c}$ superconductivity, whose low energy excitations can be described by the antiferromegnetic Heisenberg model. The Heisenberg model is with a very high accuracy realized also in the so-called spin-chain materials~\cite{spin-chain}. One-dimensional spin chains are receiving increased attention also from the cold-atoms community, where they can be experimentally realized~\cite{cold}. Particularly notorious to understand is the transport of magnetization. It has been actively studied for more than 20 years, yet the subject is still hotly debated. In particular, numerical calculations sometimes give conflicting results while there are almost no rigorous statements, exceptions are a finite Drude weight at $T=0$~\cite{Shastry:90} and recently also for $T=\infty$~\cite{Prosen:11}, both for $\Delta < 1$, signaling ballistic transport. Various, mostly numerical approaches, range from the Mazur's inequality~\cite{Zotos:97,Prosen:11}, the Bethe ansatz calculation~\cite{bethe}, exact diagonalization~\cite{exactdiag}, quantum Monte Carlo~\cite{Alvarez:02}, Lanczos method~\cite{Samir:04}, wave-packet evolution~\cite{Langer:09}, Luttinger liquid theory~\cite{Affleck:09}, master equation~\cite{JSTAT:09}, and correlation functions~\cite{Steinigeweg:09}. Because the model is solvable one would be tempted to think that it is ballistic~\cite{Castella:95}. However, recently a solvable diffusive quantum model has been found~\cite{XXdephasing}. There is also mounting evidence~\cite{JSTAT:09,Langer:09,Steinigeweg:09} that it is diffusive for $\Delta>1$.

The results presented in the present paper, together with other recent works, enable us to give a complete picture of spin transport in the linear-response regime in the Heisenberg model. For $\Delta<1$ the model is ballistic at infinite temperature~\cite{Prosen:11} as well as at any finite or zero~\cite{Shastry:90} temperature. At $\Delta=1$ our results show anomalous transport at infinite temperature. It is also plausible to expect anomalous behavior at finite temperatures. For $\Delta>1$ and infinite temperature we show diffusive transport. Recent work~\cite{Jesenko:11} shows that as one decreases temperature, the diffusion constant increases, possibly exponentially fast in $1/T$. At temperatures lower than the gap transport trivially stops. As one lowers $\Delta$ towards the isotropic point the diffusion constant diverges at any temperature, see also~\cite{Jesenko:11}.

To describe a nonequilibrium situation we couple boundary spins of the chain to magnetization reservoirs. Time evolution of the density matrix describing the chain evolves according to the Lindblad master equation,
\begin{equation}
{{\rm d}}\rho/{{\rm d}t}=\ii [ \rho,H ]+ {\cal L}^{\rm dis}(\rho)={\cal L}(\rho),
\label{eq:Lin}
\end{equation}
where the dissipative linear operator ${\cal L}^{\rm dis}$ describing bath is expressed in terms of Lindblad operators $L_k$, ${\cal L}^{\rm dis}(\rho)=\sum_k \left( [ L_k \rho,L_k^\dagger ]+[ L_k,\rho L_k^{\dagger} ] \right)$. Reservoirs are realized by two Lindblad operators at each end, $L^{\rm L}_1=\sqrt{\Gamma} \sqrt{1-\mu}\sigma^+_1, L^{\rm L}_2=\sqrt{\Gamma} {\sqrt{1+\mu}} \sigma^-_1$ at the left end and $L^{\rm R}_1 = \sqrt{\Gamma} \sqrt{1+\mu}\sigma^+_n, L^{\rm R}_2=\sqrt{\Gamma} \sqrt{1-\mu} \sigma^-_n$ at the right end, $\sigma^\pm=(\sigma^{\rm x} \pm {\rm i}\, \sigma^{\rm y})/2$. We fix $\Gamma=1$. In the diffusive regime for $\Delta >1$ the diffusion constant does not depend neither on the value of $\Gamma$ nor on the detailed Lindblad operators used to model the bath~\cite{otherL}. Our Lindblad operators are such that they always induce a nonequilibrium steady state (NESS) with almost zero average energy density, which therefore corresponds to an infinite temperature~\cite{temp}. The Hamiltonian is
\begin{equation}
H= \sum_{j=1}^{n-1} (\sigma^{\rm x}_j  \sigma_{j+1}^{\rm x} +\sigma^{\rm y}_j  \sigma_{j+1}^{\rm y})  + \Delta \sigma^{\rm z}_j  \sigma_{j+1}^{\rm z}.
\label{eq:Hzz}
\end{equation}

{\bf \em Small $\Delta$--} In this section we are interested in small $\Delta$, so that the $\sigma^{\rm z}_j  \sigma_{j+1}^{\rm z}$ term acts as a perturbation of the XX model. Analytical perturbative results will serve as a spring-board for the discussion of transport for not so small values of $\Delta<1$. We split the Lindblad superoperator into two parts, ${\cal L}={\cal L}^{(0)}+\Delta\,{\cal L}^{(\rm zz)}$, where ${\cal L}^{(0)}$ is the part for $\Delta=0$, while ${\cal L}^{(\rm zz)}$ is the perturbation, i.e., ${\cal L}^{(\rm zz)}(\rho)= \ii [\rho, \sum_j \sigma^{\rm z}_j  \sigma_{j+1}^{\rm z} ]$. For $\Delta=0$ the NESS solution of this master equation, denoted by $\rho_0$, is nondegenerate and ballistic and can be neatly written in a matrix product operator form with matrices of fixed dimension $4$~\cite{JPA10}. Perturbation series ansatz for the NESS for small $\Delta$ is $\rho=\rho_0+\Delta\cdot \rho_1+\Delta^2\cdot \rho_2+\cdots$. We want to calculate the first two orders in $\Delta$. Also, in all terms we will be interested only in the lowest order terms in the driving $\mu$, meaning that we study linear-response behavior. Plugging this ansatz into NESS equation ${\cal L}(\rho)=0$, using ${\cal L}^{(0)}(\rho_0)=0$, and then equating terms with the same order in $\Delta$, gets us two matrix equations, ${\cal L}^{(0)}(\rho_1)=-{\cal L}^{(\rm zz)}(\rho_0)$ and ${\cal L}^{(0)}(\rho_2)=-{\cal L}^{(\rm zz)}(\rho_1)$. A known zeroth-order solution $\rho_0$ can be used to get inhomogeneous terms in the linear equations for $\rho_1$, which can in turn be used to get $\rho_2$. The only problem is that the number of linear equations is exponentially large in the length of the chain $n$. Nevertheless, several general remarks can be made because we know $\rho_0$ (it contains only $\sigma_{1,n}^{\rm z}$, spin current $j_k=2(\sigma_k^{\rm x} \sigma_{k+1}^{\rm y}-\sigma_k^{\rm y}\sigma_{k+1}^{\rm x})$, and their products) and the action of ${\cal L}^{(0)}$ and ${\cal L}^{(\rm zz)}$: the first order term $\rho_1$ does not contain any magnetization $\sigma_j^{\rm z}$ or spin current $j_k$; they appear only in the 2nd order term $\rho_2$. Therefore, to the lowest order in perturbation $\Delta$ the current and the magnetization profile do not change! They are the same as for the ballistic XX model at $\Delta=0$. Because the perturbation of a nondegenerate NESS is nonsingular, fixing $n$, one can always find a sufficiently small $\Delta$ such that the perturbative expansion will converge and the system is ballistic. However, for the transport in the thermodynamic limit the relevant order of limits is first fixing $\Delta$ and $\mu$ and only then sending $n\to \infty$. This limit is more difficult to treat because higher order terms, for instance $\rho_2$, can grow with $n$ faster than the first order term $\rho_1$, causing the convergence radius to shrink as $n \to \infty$. As we will see, this is indeed what happens.

We have found the exact expressions for $\rho_1$ and $\rho_2$ for small $n\le 7$; the coefficients in front of all the terms are rational functions with a rather large denominators and as such not very transparent. We therefore do not give their precise form here but rather focus on their scaling with $n$ in order to infer the convergence radius of the perturbative expansion. 
Because we are interested in the spin transport, we need to know the behavior of the correction in $\rho_2$ that involves $\sigma_j^{\rm z}$ and spin current $j_k$. It turns out that the correction in the current is the same for all sites while the correction to the magnetization depends on the position. For the spin current, the coefficient in front of the term $j_k/2^n$ is , $-\mu \frac{67}{404}$ for $n=4$, $-\mu \frac{69235}{513248}$ for $n=5$, while it is $-\mu \frac{45569624481}{243264258368}$ for $n=6$ and $-\mu\frac{56317144998719121983}{117362105703777609136}$ for $n=7$. If one looks at the dependence of these coefficients on $n$ one notices that it is to a very good approximation linear. 
Fitting gives the dependence $c(n)=0.4286\,(n-2.436)$, with deviations being possibly exponentially small in $n$. Similar corrections, all growing linearly with $n$ are also found for $\sigma_j^{\rm z}$ (the one for $\sigma_1^{\rm z}$ is in fact equal to $-2$ times the one for the current, for other spins prefactors are larger). To sum up, the expectation value of the spin current is to order $\Delta^2$ (and to linear order in $\mu$) equal to $\ave{j_k}\approx \mu-\mu \Delta^2\,\, 0.429\,(n-2.436)$.
Because the 2nd order correction grows with the system size, the perturbative expansion holds only for $\Delta \le 1/\sqrt{n}$, and therefore breaks down in the  thermodynamic limit.
Unfortunately, from our analytical calculation one therefore can not decide about the nature of the spin transport at finite $\Delta$~\cite{foot1}.

\begin{figure}[t!]
\centerline{\includegraphics[width=0.5\textwidth]{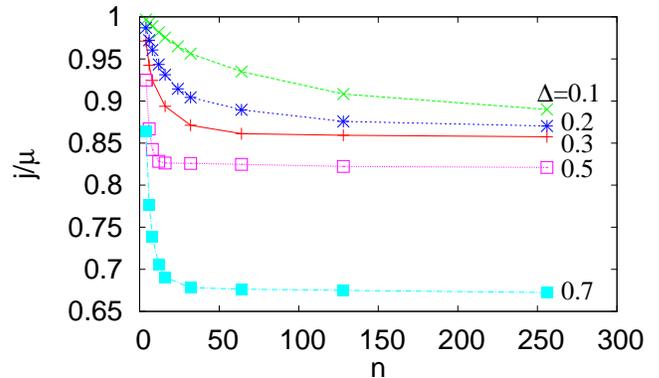}}
\caption{Expectation value of the spin current on system size $n$ obtained by tDMRG. After sufficiently large $n_{\rm c}$ the current converges to a $n$-independent value, signaling a ballistic spin transport. For small $\Delta$ this happens at $n_{\rm c} \sim 1/\Delta^2$, at larger the scaling seems to be different.}
\label{fig:jodn}
\end{figure}

For larger $\Delta$ we used time-dependent density matrix renormalization (tDMRG) simulations~\cite{JSTAT:09} with $\mu=0.02$ to get expectations of magnetization and spin current in the NESS. Results are in Fig.~\ref{fig:jodn}.
For $\Delta < 1$ the current saturates for sufficiently large $n$, with the saturation current monotonically decreasing with $\Delta$. The system is therefore ballistic. For small $\Delta < 0.5$ the characteristic $n_{\rm c}$ at which $j$ converges to a constant value scales as $\sim 1/\Delta^2$, which is the same as the scaling of the analytical perturbative result. Even though perturbation theory fails as $n \to \infty$, the scaling $\Delta^2 n = {\rm const.}$ apparently caries over beyond the perturbative result.

\begin{figure}[h!]
\centerline{\includegraphics[width=0.25\textwidth]{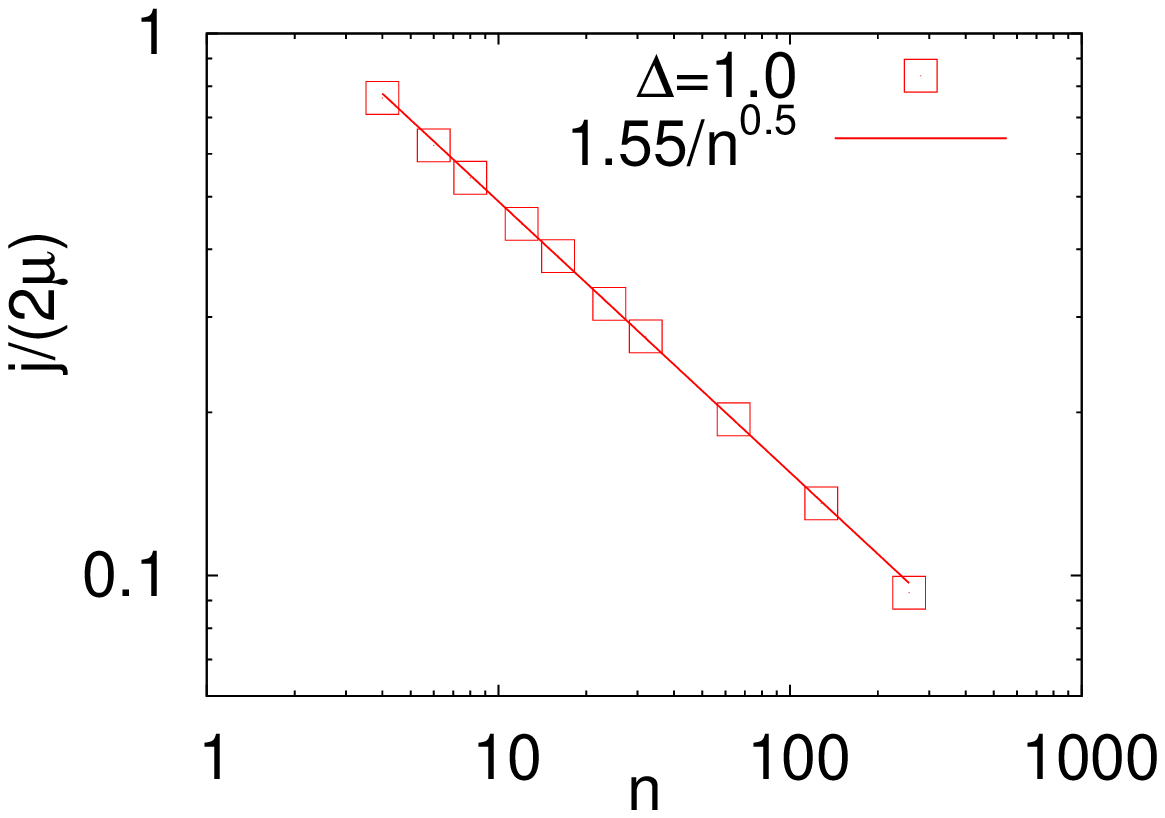}\hskip-0mm\includegraphics[width=0.25\textwidth]{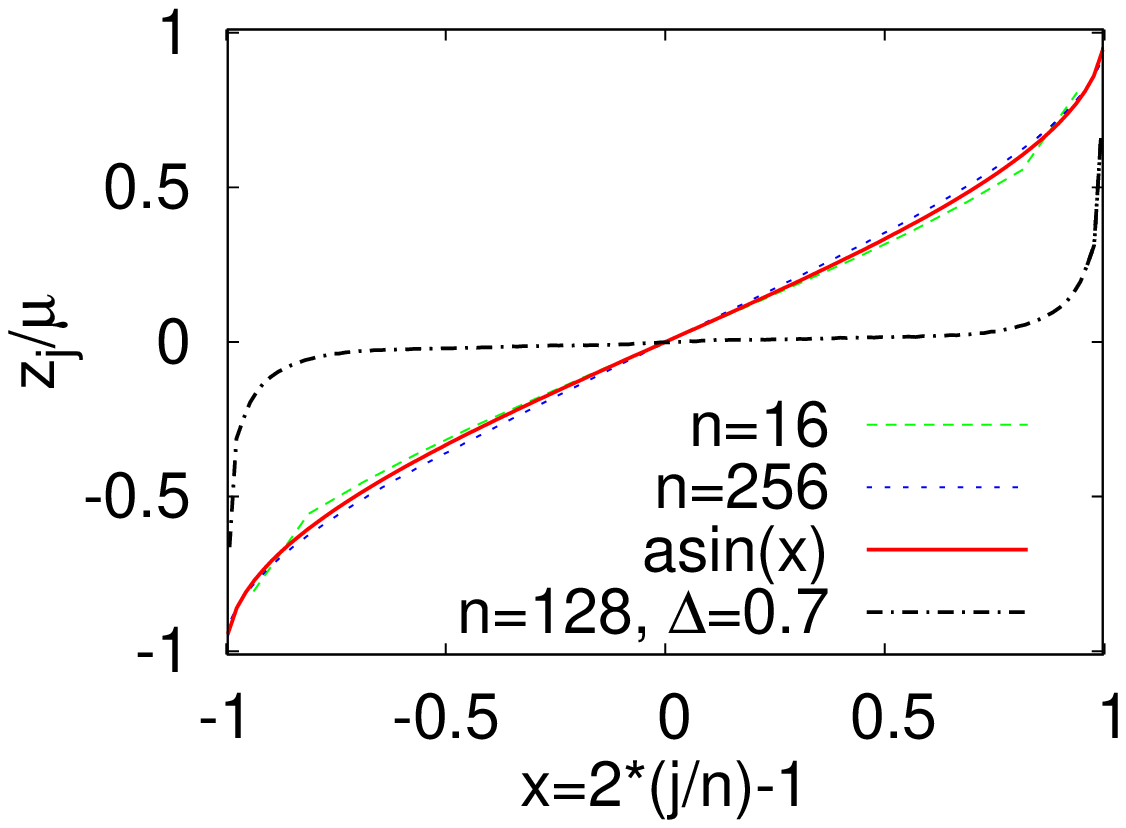}}
\caption{Left frame: the scaling of the spin current on the system size $n$ for $\Delta=1$. The current decays only as $\sim 1/\sqrt{n}$ (solid line), indicating a superdiffusive transport. Right frame: Scaling of the magnetization profile at $\Delta=1$ (two overlapping dashed curves) is very similar to $\arcsin{x}$ (red solid curve). For $\Delta <1$ the profile is flat (dot-dashed curve).}
\label{fig:D1}
\end{figure}

{\bf \em Isotropic--} At the isotropic point the spin current at fixed driving scales as $j \sim 1/\sqrt{n}$, nicely seen in tDMRG data in Fig.~\ref{fig:D1}. The isotropic Heisenberg model at infinite temperature therefore display anomalous diffusion, with the diffusion constant diverging as $D \sim \sqrt{n}$. This is the first observation of an anomalous diffusion in a coherent (Hamiltonian) quantum system.
Furthermore, the magnetization profile along the chain has a nice scaling with $n$ and $\mu$. As can be seen in Fig.~\ref{fig:D1}, the dimensionless scaling function looks to be very close to $\arcsin{x}$, however, deviations seen in the figure seem to be larger than the finite-size or numerical accuracy effects.

\begin{figure}[ht!]
\centerline{\includegraphics[width=0.5\textwidth]{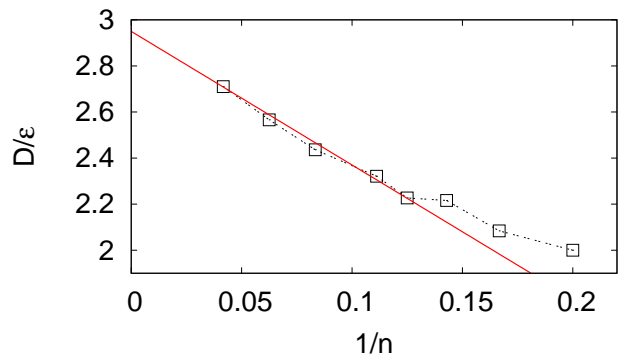}}
\caption{Finite-size scaling of the diffusion constant for large $\Delta$. Finite value for $n \to \infty$ indicates diffusion.}
\label{fig:ratio}
\end{figure}

{\em \bf Large $\Delta$--} Perhaps the most interesting regime is for $\Delta>1$ where numerical calculations point to a diffusive transport at infinite temperature~\cite{JSTAT:09,Steinigeweg:09}. Because analytical treatment seems to be hard, we choose to study the case of large $\Delta$, where the Hamiltonian is close to the Ising one. The case of large $\Delta$ can be equivalently reformulated with the Hamiltonian $H= \sum_{j=1}^{n-1} \epsilon (\sigma^{\rm x}_j  \sigma_{j+1}^{\rm x} +\sigma^{\rm y}_j  \sigma_{j+1}^{\rm y})  + \sigma^{\rm z}_j  \sigma_{j+1}^{\rm z}$, with $\epsilon=1/\Delta$~\cite{foot2}. The NESS for $\epsilon=0$ is exponentially degenerate. It is easy to see that the eigenvector of a dissipative bath part is ${\cal L}^{\rm bath}_1 (\mathbbm{1}-\mu \sigma_1^{\rm z})=0$ at the left end and ${\cal L}^{\rm bath}_n (\mathbbm{1}+\mu \sigma_n^{\rm z})=0$ at the last spin. Because of $[\sigma_j^{\rm z},H_{\epsilon=0}]=0$, any operator of the form $(\mathbbm{1}-\mu \sigma_1^{\rm z})\otimes x \otimes (\mathbbm{1}+\mu \sigma_n^{\rm z})$, with $x$ being an arbitrary combination of $\mathbbm{1}_j$ and $\sigma_j^{\rm z}$, is a zero eigenstate of the Lindblad superoperator, i.e., the NESS. There are $2^{n-2}$ independent states of this form. Besides these, there are additional NESS states, namely, those for which ${\cal L}^{(\epsilon=0)}(x)=0$ holds and where $x$ now includes also $\sigma_j^{\rm x,y}$. This increases the degeneracy even further. Because of this high degeneracy perturbation theory is more difficult than for small $\Delta$. An important thing to note is that the NESS for such Ising chain can support an arbitrary magnetization profile, while the spin current is always zero. The Ising spin chain is therefore a perfect insulator. Exponentially high degeneracy can now be understood also as being due to the isolation of the bulk from the boundaries, so that spins in the bulk ``do not know'' about the reservoirs at the boundaries. High degeneracy is therefore generic and can not be removed by a different choice of Lindblad operators. Perturbation $\epsilon$ breaks this high degeneracy making the NESS nondegenerate. The gap between two eigenvalues of the Lindblad superoperator with the largest real parts scales as $\epsilon^2$ for small perturbations. This means that if we want to calculate the lowest order corrections in the NESS exactly, we have to expand it upto order $\epsilon^2$. Perturbative expansion can be for small $\mu$ written as $\rho=\frac{1}{2^n}(\mathbbm{1}+\mu \rho_0+\mu \epsilon \rho_1+\mu \epsilon^2 \rho_2+\cdots)$. The resulting linear equations for unknown $\rho_0, \rho_1$ and $\rho_2$ are ${\cal L}^{(0)}(\rho_0)=0, {\cal L}^{(0)}(\rho_1)+{\cal L}^{(\rm xx)}(\rho_0)=0$, and ${\cal L}^{(0)}(\rho_2)+{\cal L}^{({\rm xx})}(\rho_1)=0$, where we have split the superoperator into an unperturbed part ${\cal L}^{(0)}$ and the perturbation ${\cal L}^{(\rm xx)}(\rho)= \ii [\rho, \sum_j \sigma^{\rm x}_j  \sigma_{j+1}^{\rm x}+ \sigma^{\rm y}_j  \sigma_{j+1}^{\rm y}]$. 
Using the appropriate ansatz, we have obtained exact solutions for small $n\le 9$, however, they are again complicated, involving many terms. We only point out features important for the spin transport. The first observation is that $\rho_0$ can contain only terms that are already present in the NESS for $\epsilon=0$. This includes magnetization, but not the spin current. Therefore, operators $\sigma_j^{\rm z}$ will be present in $\rho_0$, while the spin current will be present only in the first order term $\rho_1$ (because ${\cal L}^{(\rm xx)}(\sigma_j^{\rm z})$ will result in the current). Spin current is therefore always proportional to $\epsilon$ ($=1/\Delta$), while the magnetization scales as $\epsilon^0$. Writing out the equation involving the coefficient $c$ in front of the spin current $j_k$ in $\rho_1$, one gets $2c=z_1-z_2-(h_{1,2}z_3)$, where $z_k$ is the coefficient in front of $\sigma_k^{\rm z}$ in $\rho_0$ and $(h_{1,2}z_3)$ is the coefficient in front of $(\sigma_1^{\rm x}\sigma_2^{\rm x}+\sigma_1^{\rm y}\sigma_2^{\rm y})\sigma_3^{\rm z}$ in $\rho_1$. Our exact analytical solutions for small $n$'s show that the term $(h_{1,2}z_3)$ is always equal to $2c$ for $n\ge 4$. Therefore, for large $n$ one has an exact relation $4c=z_1-z_2$. This states that if the magnetization profile is linear on average, then the spin current scales as $\sim 1/n$ and the transport is diffusive. Of course, showing that $z_1-z_2 \sim 1/n$ might be no easier than showing this for the current. Exact solutions give the expectation value of the current $j=\mu \epsilon\, {\rm tr\,}{(j_k\rho_1)}/2^n$ as $(n-1)\cdot j/(2\mu \epsilon)=3, \frac{5}{2}, 2, \frac{25}{12}, \frac{195}{88}, \frac{225127}{101088}$, for $n=3,\ldots,8$. To access the limiting value we have looked at the convergence of $j/\epsilon \nabla \mu=(n-1)(z_1-z_2)$ with $n$. If the transport is diffusive, this coefficient should converge to the diffusion constant $D$. In the Fig.~\ref{fig:ratio} we plot the values of these exact coefficients, together with numerically obtained ones for $n\le 24$. The scaling seems to be linear in $1/n$ enabling us to obtain the limiting value of the coefficient as $n \to \infty$. 
Using this limit we can predict that the spin current goes as $j \asymp \epsilon 2\mu \frac{2.95}{n}+{\cal O}(\mu\epsilon^3)$, resulting in the diffusion constant~\cite{foot3}
\begin{equation}
D \asymp 2.95/\Delta.
\label{eq:D}
\end{equation}
The dependence of $D$ on $\Delta$ has been discussed in~\cite{Stein:10}. The 2nd order term $\rho_2$ (as well as the 3rd) does not contain any corrections to the current. From the analysis of the solutions for small $n$ it also does not appear that they would grow with $n$. The convergence radius of the perturbative series is therefore finite and does not decrease as $n\to\infty$.

\begin{figure}[t!]
\centerline{\includegraphics[width=0.5\textwidth]{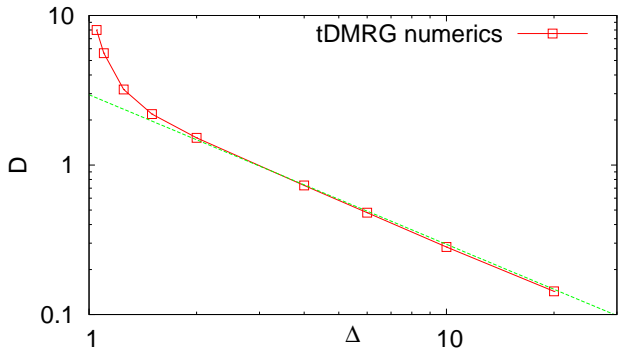}}
\caption{The dependence of the diffusion constant $D$ on $\Delta$. Straight line is the theoretical prediction (\ref{eq:D}) obtained from the perturbative treatment of the Ising model.}
\label{fig:kappaP}
\end{figure}

To verify the theoretical prediction for a diffusion constant (\ref{eq:D}) we have again performed tDMRG simulations for a range of $\Delta$ as well as $n$. For each $\Delta$ a diffusive scaling of the current $j \sim 1/n$ has been checked and the prefactor, being the diffusion constant, determined. The results are plotted in Fig.\ref{fig:kappaP}, together with Eq.(\ref{eq:D}). 
We can see a perfect agreement for large $\Delta$, with the perturbative result (\ref{eq:D}) holding upto quite small $\Delta$~\cite{foot4}.

{\em Conclusion--} By using perturbation theory in the limit of small and large anisotropies $\Delta$ as well as large-scale numerical simulations we have shown that for $\Delta<1$ the Heisenberg model displays ballistic spin transport. At the isotropic point transport is anomalous, with the current scaling as $\sim 1/\sqrt{n}$. This is the first observation of an anomalous transport in a coherent quantum system and has strong implications for an unexplained high heat conductivity measured in spin-chain materials~\cite{exper}. For $\Delta>1$ we show that the transport is diffusive and inversely proportional to $\Delta$ for large anisotropies. Support by the Program P1-0044 and the Grant J1-2208 of the Slovenian Research Agency, the project 57334 by CONACyT, Mexico, and IN114310 by UNAM is acknowledged.


\begin{thebibliography}{}
\bibitem{otherL} We have checked that a 2-spin bath involving 16 Lindblad operators on the boundary two spins, and used before in~\cite{JSTAT:09}, leads to essentially the same results at $\Delta=1$ and $\Delta=1.5$. For $\Delta<1$ the influence of Lindblad operators is more delicate and we do not discuss it here.
\bibitem{foot1} In order to make sure that there are no faster growing terms in 3rd and higher orders we verified the perturbative prediction by tDMRG simulations, by which we can treat larger systems. It is found that the perturbative result indeed breaks down when $\Delta^2 c(n) \approx 0.1$.
\bibitem{foot2} The NESS in such formulation is exactly the same as for the Hamiltonian $\sum_{j=1}^{n-1} \sigma^{\rm x}_j  \sigma_{j+1}^{\rm x} +\sigma^{\rm y}_j  \sigma_{j+1}^{\rm y}  + \Delta \sigma^{\rm z}_j  \sigma_{j+1}^{\rm z}$ with the coupling strength $\Gamma=\Delta$ (the coupling to reservoars should scale the same as the largest parameter in $H$).
\bibitem{foot4} A small discrepancy at largest $\Delta$ can be attributed to finite-size effects as the numerical simulations get more difficult there and we could simulate only chains of order $n\sim 40$, whereas we could go upto $n \sim 128$ at $\Delta=2.0$.
\end{thebibliography}
\end{document}